\documentclass[final,times,5p,twocolumn]{elsarticle}
\usepackage{epsfig}
\usepackage{graphicx}
\usepackage{dcolumn}
\usepackage{bm}
\usepackage{graphicx}
\usepackage{dcolumn}
\usepackage{bm}
\def\br{\begin{eqnarray}}
\def\er{\end{eqnarray}}
\def\be{\begin{equation}}
\def\ee{\end{equation}}

\def\({\left(}
\def\){\right)}

\def\<{\left\langle}
\def\>{\right\rangle}

\begin{document}

\title{Scalar coupling evolution in a non-perturbative QCD resummation scheme}
 
\author{J.~D.~Gomez$^1$ and A.~A.~Natale$^{1,2}$}
\ead{jgomez@ufabc.edu.br,natale@ift.unesp.br}
\address{$^1$Centro de Ci\^encias Naturais e Humanas, Universidade Federal do ABC, 09210-170, Santo Andr\'e - SP, Brazil \\
$^2$Instituto de F\'{\i}sica Te\'orica, UNESP, Rua Dr. Bento T. Ferraz, 271, Bloco II, 01140-070, S\~ao Paulo - SP, Brazil}

\begin{abstract}
We compute the Standard Model scalar coupling ($\lambda$) evolution in a particular QCD resummation scheme, where the QCD coupling becomes infrared finite due to the
presence of a dynamically generated gluon mass, leading to the existence of a non-perturbative infrared fixed point. We discuss how this scheme can be 
fixed taking recourse to phenomenological considerations in the infrared region. The QCD $\beta$ function associated to this non-perturbative
coupling when introduced into the SM renormalization group equations increases the $\lambda$ values at high energies.   
\end{abstract}



\maketitle

One of the techniques to study QCD at low energies are the Schwinger-Dyson Equations (SDE). This is an analytic method but with the complication of dealing with an infinite
tower of coupled integral equations. Because of this, when using the SDE, one have to make a truncation in the system of equations, which has to be done in a way that the
symmetries of the theory are not broken, in particular, a rough truncation in the SDE can cause a violation of the gauge symmetry. In the recent years an enormous progress
has been made in solving SDE in a gauge invariant way using the so called Pinch Technique \cite{pinch}, as a result it has been found the existence of a dynamical
gluon mass in the propagator of the gluon field, as suggested many years ago by Cornwall \cite{cornwall}. Another non-perturbative technique is Lattice QCD \cite{green}, which
implies heavy numerical calculations requiring a considerably high amount of computer power. Results obtained in QCD lattice simulations are in agreement with the SDE in what concerns
dynamical gluon mass generation \cite{aguilar,binosi}. 

The infrared QCD coupling turns out to be infrared finite when gluons develop a dynamically generated mass. This point was already emphasized
in Ref.\cite{cornwall}; was also discussed at length in Ref.\cite{papa1,papa2}, and leads to an infrared fixed point, which is a property of dynamical mass generation in non-Abelian theories \cite{us1}. The phenomenological consequences of such infrared finite coupling, or non-perturbative
fixed point, have been discussed in Ref.\cite{us2}, and recently we have discussed how this non-perturbative fixed point can change the local minimum of a renormalization group improved effective potential \cite{us3}. This change of minimum state may produce noticeable 
modifications in the physical properties of the model studied in Ref.\cite{us3}.

In view of the results of Ref.\cite{us3} we will study the effect of the non-perturbative fixed point present in QCD with dynamically
massive gluons in the scalar coupling evolution of the Standard Model (SM). It should be noticed that we shall be dealing with a very
particular QCD resummation scheme, provided by the results of the Pinch Technique applied to the SDE, which have been argued that
can lead to off-shell Green's functions that are locally gauge invariant and renormalization group invariant \cite{cornwall2}.
The QCD $\alpha_s$ coupling that we shall consider will not depend on the renormalization point $\mu$ but on the dynamical gluon
mass $m_g (k^2)$ and, of course, on the QCD characteristic scale $\Lambda_{QCD}\equiv \Lambda$. This coupling contains the
effect of summation of several loops according to the calculations detailed in Refs.\cite{pinch,cornwall,papa1,papa2}, and the free
parameter in this particular scheme, namely the infrared value of the dynamical gluon mass, will be fixed by taking recourse
to phenomenological considerations about the coupling constant IR behavior. 

The first calculation of the IR frozen QCD coupling in the presence of a dynamically generated gluon mass was
obtained in Ref.\cite{cornwall}, leading to the following coupling:
\be
g^2(k^2)=\frac{1}{\beta_0 \ln\Big[\frac{k^2+4 \cdot m_{g}^{2}}{\Lambda_{QCD}^2}\Big]}=4\pi\alpha_s (k^2),
\label{eq:1}
\ee
where $\beta_0=(11N-2n_q)/48\pi^2$ with $n_q$ quark flavors and $N=3$. $m_g$ is the IR value of the dynamical gluon mass
$m_g(k^2)$, which naturally goes to zero at high energies.  Note that the running charge has been obtained as a fit to the SDE solutions
in the pure gluon theory, but quarks are introduced at leading order just by its effect in the first $\beta$ function coefficient. 
The most important factor in the $\alpha_s (0)$ frozen IR value of the QCD coupling determination
is the ratio $m_g/\Lambda_{QCD}$, which was approximately determined to be a factor of ${\cal{O}}(2)$ \cite{cornwall}. 

Another possible, and more detailed, fit of the effective QCD charge determined from the SDE solutions within the pinch technique 
is given by \cite{agui2}
\be
\alpha_s (k^2)=\Bigg[4\pi {\beta_0}\ln\Bigg(\frac{k^2+f\big(k^2,m^2(k^2)\big)}{\Lambda_{QCD}^{2}}\Bigg)\Bigg]^{-1},
\label{eq:2}
\ee
where the function $f\big(k^2,m^2(k^2)\big)$ is given by
\be
f\big(k^2,m^2(k^2)\big)=\rho_1 m^2(k^2)+\rho_2\frac{m^4(k^2)}{k^2+m^2(k^2)} \, ,
\label{eq:3}
\ee
where $\rho_1,\, \rho_2 $ are fitting parameters and the function $m^2(k^2)$ represents a running dynamical gluon mass that, apart from negligible logarithmic factors, is approximately given by \cite{agui2,lav,aa}
\be
m_g^2(k^2)\approx \frac{m_{g}^{4}}{k^2+m_{g}^{2}},
\label{eq:4}
\ee
where $m_{g}$ is the effective dynamical gluon mass IR value. For $m_g = 500$ MeV and $\Lambda_{QCD}=300$ MeV, the
best fit for the running charge gives $\rho_1 =4.5$ and $\rho_2=-2$ \cite{agui2}. 

Eqs.(\ref{eq:2}) and (\ref{eq:3}) are just one fit for one
specific $m_g$ and $\Lambda_{QCD}$ set of values, and it should be numerically calculated each time that we vary one of these parameters, 
while Eq.(\ref{eq:1}) can be used in the case of 
different $m_g$ values. As a crude approximation, without the need of calculating the full SDE, we could just replace $m_g^2$ in 
Eq.(\ref{eq:1}) by $m_g^2(k^2)$ given by Eq.(\ref{eq:4}). When we perform this replacement and compute the coupling for $m_g= 500$ MeV and 
$\Lambda_{QCD}=300$ MeV we obtain a coupling that differs slightly from the coupling shown in Eq.(\ref{eq:2}) in the region of $k^2\approx 2 m_g^2$. 
Another approach can be obtained just assuming the following simple
fit for the coupling constant
\br
4\pi\alpha_s (k^2) &&\approx  \frac{1}{\beta_0 \ln\Big[\frac{4 \cdot m_{g}^{2}}{\Lambda_{QCD}^2}\Big]} \theta (1 {\textit{GeV}}^2 - k^2) \nonumber \\
&&+ \frac{\kappa}{\beta_0 \ln\Big[\frac{k^2+4 \cdot m_{g}^{2}/k^2}{\Lambda_{QCD}^2}\Big]}\theta (k^2 -1 {\textit{GeV}}^2),
\label{eq:5}
\er
where $\kappa$ is a constant that provides the interpolation between the constant behavior of the IR coupling with its high energy behavior, where
the dynamical gluon mass falloff as $1/k^2$ \cite{lav}. In any case it is important to stress that the non-perturbative QCD coupling associated to
the phenomenon of dynamical gluon mass generation matches exactly with the perturbative one at high energies.

Exactly as performed in Ref.\cite{us3}, where the non-perturbative behavior of the coupling constant was used in the calculation of a renormalization group improved
effective potential, we would like to use the coupling constant discussed above to compute the SM scalar coupling evolution. It is clear that such coupling
appears in a non-perturbative QCD resummation scheme, and we shall need to take recourse of phenomenological considerations to fix its IR behavior before considering its
use in the renormalization group (RG) equations. Of course, as a requirement for stability of the calculation, the IR non-perturbative QCD coupling surely cannot be large, or at least it must be smaller or of the order of the top quark Yukawa coupling, since we are going to confront it with other SM couplings. However this is
exactly what is going to happen, and most of the phenomenological models trying to extract the $\alpha_s$ IR coupling value seems to 
indicate a small number. For instance, the description of jet shapes observables
require $\alpha_s (0)$ to be of the order $0.63$ \cite{Webber}, the famous models of quarkonium potential calculations of Ref.\cite{quarko} use an IR coupling of order 
$\alpha_s(0) \approx 0.6$,
the ratio $R_{e^+e^-}$ computed by Mattingly and Stevenson \cite{Matt} can fit the data with 
$\alpha_s (0)/\pi \approx 0.26$, analysis of $e^+e^-$
annihilation, as well as bottomonium and charmonium fine structure in the framework of the background perturbation theory may lead to a
frozen value of the coupling constant as low as $\alpha_s (0) \approx 0.4$ \cite{dezset}. 
Recent analysis of experimental data on the unpolarized structure function of the proton indicates that \cite{court}:
\be
0.13 \leq {\alpha_{s,NLO} (scale \rightarrow 0)/\pi} \leq 0.18 \,\, ,
\label{eq:6}
\ee
what is also consistent with $\alpha_s$ values extracted from the GDM sum rule \cite{court}.

Other phenomenological calculations considering
a finite IR QCD coupling can be found in Ref.\cite{sev}, 
and a compilation of some results can be seen in Ref.\cite{nat}. We can add to these phenomenological computations the theoretical 
$\alpha_s (0)$ value obtained through the functional Schr\"odinger equation which is equal to $0.5$ \cite{co1}, and the quite extensive list of an IR finite effective coupling calculations based on the Schwinger-Dyson equations (SDE) within the Pinch Technique 
\cite{pinch,cornwall,cornwall3}, which also lead to a successful strong interaction phenomenology \cite{us2,natale1}.
 
In the scheme that we  are following the QCD coupling freezes in the IR at one reasonably small value, and, as shown in Eq.(\ref{eq:1}) and Eq.(\ref{eq:2}), the IR value is directly related to the IR value of the dynamical gluon mass, which is 
approximately bounded to $m_g > (0.6-1.2) \Lambda_{QCD} $ in 
order to ensure that there be no tachyons in the gluon propagator \cite{corn2}. The preferred phenomenological value 
of this mass is associated to the ratio $m_g/\Lambda_{QCD}\approx 2$ \cite{cornwall,us2}. For a dynamical gluon mass
$m_g  = 1.2 \Lambda_{QCD}$ we obtain an IR QCD coupling of
order of $0.8$. Therefore, according to the many phenomenological determinations of the $\alpha_s (0)$ 
values described previously, and with the phenomenological determinations of the effective dynamical gluon mass \cite{us2}
we will consider the following range for the IR value of this coupling
\be
0.4 \leq {\alpha_{s} (0)} \leq 0.8 \,\, ,
\label{eq:7}
\ee  
what is in agreement with previous discussions about the IR value of the strong coupling \cite{nat}. Note that, according to Eq.(\ref{eq:1}), the possible 
values of the ratio $m_g/\Lambda_{QCD}$ for two quark flavors, to be in agreement with Eq.(\ref{eq:7}), are in the range:
\be
1.2 \leq \frac{m_g}{\Lambda_{QCD}} \leq 2.86 \,\, .
\label{eq:8}
\ee
Which are also in agreement with the many different determinations of $\alpha_s(0)$ and $m_g$ that can be found in the literature. At this point
we have set up the parameters necessary in this QCD resummation scheme to calculate the SM scalar self coupling evolution.

The evolution of the scalar self coupling $\lambda$ appearing in the SM,
where the scalar field may acquire a vacuum expectation value $v\approx 246.2$ GeV, are giving by the standard RG equations where each SM ordinary coupling $\alpha_i$ is governed
by the respective $\beta$ function
\be
\beta_i (\alpha_i) = \mu^2 \frac{d}{d\mu^2} \alpha_i (\mu),
\label{eq:9}
\ee
where $\alpha_i$ represents $\lambda$ and any gauge or Yukawa SM couplings. The SM stability up to the Planck scale requires 
$\lambda \geq 0$, and the evolution 
of this coupling is determined solving the coupled system of differential equations given by Eq.(\ref{eq:9}).

The SM RG equations with the one loop $\beta$ functions in the ($\overline{\mathrm{MS}}$) scheme (up to Eq.(13)) are given by
\br
 \beta_\lambda && = \frac{1}{(4\pi)^2}\Big[24\lambda^2-6y_{t}^4 + \frac{3}{8}\big(2g_{2}^4+(g_{2}^2+g_{1}^2)^2\big) \nonumber \\
               &&-(9g_{2}^2+3g_{1}^2-12y_{t}^{2})\lambda\Big], \\
 \beta_{y_t}   && = \frac{y_{t}}{(4\pi)^2}\Big[-\frac{9}{4}g_{2}^{2}-\frac{17}{12}g_{1}^{2}-8g_{3}^{2}+\frac{9}{2}y_{t}^{2}\Big],  \\
 \beta_{g_1 }  && = \frac{1}{(4\pi)^2}\frac{41}{6}g_{1}^3,  \\
 \beta_{g_2}   && = \frac{1}{(4\pi)^2}\frac{-19}{6}g_{2}^3, \\
 \beta_{g_3}   && = -\beta_0 g^3 \frac{e^t}{e^t + 4\frac{m_{g}^{2}(t)}{\Lambda^2}}\Big(1-\frac{4}{e^t + \frac{m_{g}^{2}}{\Lambda^2}}\frac{m_{g}^{2}(t)}{\Lambda^2}\Big),
\label{eq:x}
\er
where $t=\log\frac{k^2}{\Lambda^2}$, $\frac{m_g^2 (t)}{\Lambda^2} = \frac{(m_g^4/\Lambda^4)}{[e^t+m_g^2/\Lambda^2]}$, and $\beta_0=\frac{11N-2n_q}{48\pi^2}$
is the first coefficient of the QCD $\beta$ function. Note that $\beta_\lambda$, $\beta_{y_t}$, $\beta_{g_1 }$, $ \beta_{g_2}$ and $ \beta_{g_3}$ are respectively
the scalar, Yukawa top quark, $U(1)$, $SU(2)$ and $SU(3)$ $\beta$ functions. However $\beta_{g_3}$ 
has been changed by the non-perturbative QCD $\beta$ function generated by the non-perturbative coupling described previously. This is the approach 
pursued in Ref.\cite{us3}, and also in Ref.\cite{wet} in a context that includes gravity in the calculation. 

\begin{figure}[h]
\begin{centering}
\hspace{-0.3in}
\includegraphics[scale=0.35]{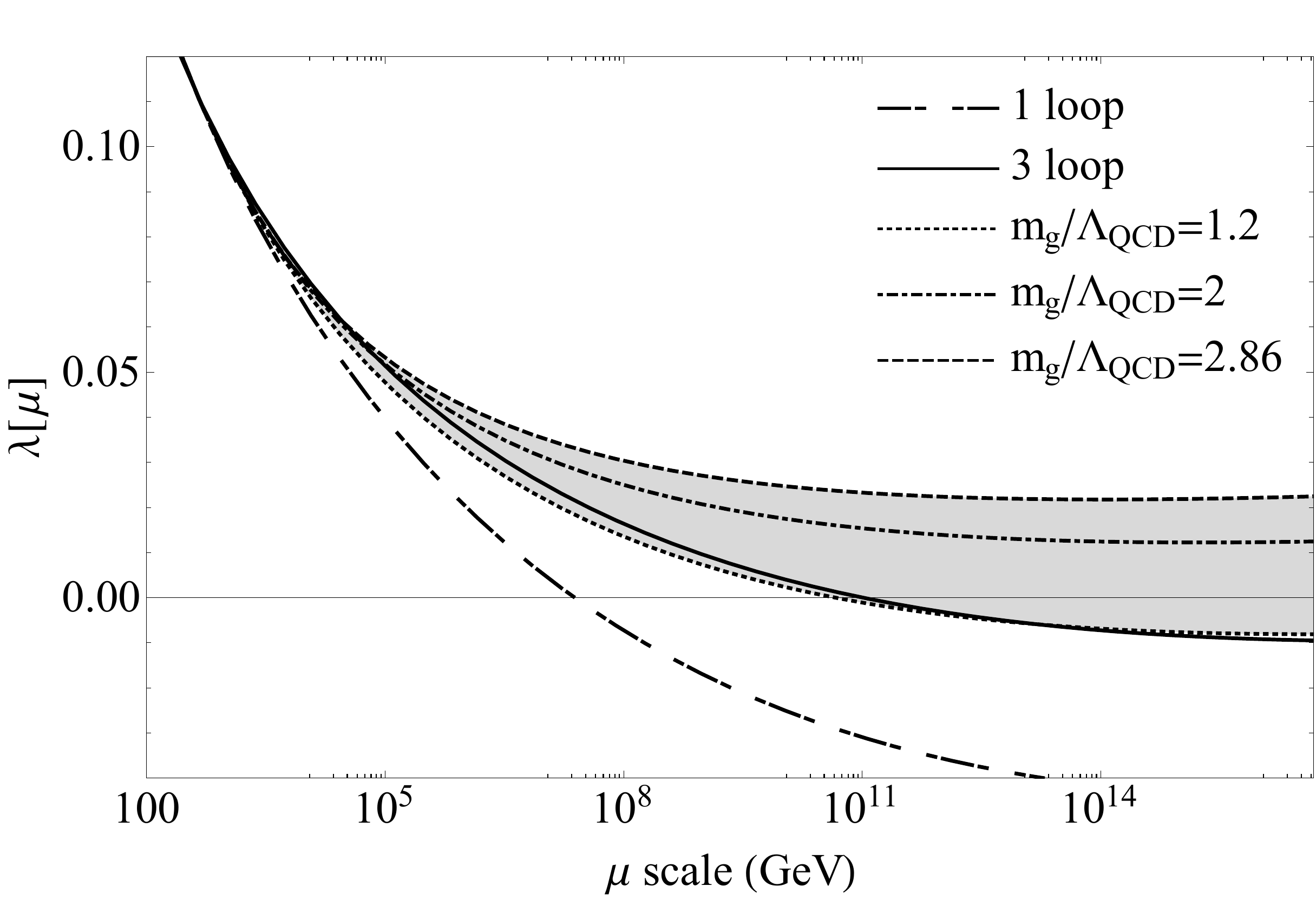} 
\par\end{centering}
\caption{Scalar coupling ($\lambda$) evolution. The $1$ and $3$-loops standard pertubative calculations are shown by the dashed and continuous lines. The shaded area is obtained varying the $m_g/\Lambda_{QCD}$ values between $m_g/\Lambda_{QCD}\approx 1.2$ and $2.86$ (corresponding to $\alpha_s (0)$ between $0.8$ and $0.4$). The curve in the middle of the shaded area correspond to
the $\lambda$ evolution for the phenomenologically preferred ratio of $m_g/\Lambda_{QCD}=2$, and this curve do not cross the line $\lambda = 0$ up to the Planck scale.}
\label{fig1} 
\end{figure}

To solve the RG equations we will use the same one-loop (and three-loop) SM $\beta$ functions used in Ref.\cite{zol1,zol2}. In this case our results
are shown in Fig.(\ref{fig1}) indicated by $1$ and $3$-loops (respectively small and large dashed curve and the continuous one), which agree with the ones of Ref.\cite{zol1,zol2} and allow us to check the numerical code. We have used exactly the same initial conditions shown in Table 1 of Ref.\cite{zol2} at $\mu=M_t$, where the top mass is $M_t = 172.9\pm0.6\pm0.9$ GeV,
$M_H = 125.7$ GeV, and 
\be
\alpha_s (M_Z)=0.1185\pm0.0007 \,\, .
\label{eq:77}
\ee
For the $\lambda$ evolution at very high energies we assumed $n_q =6$, and no particular attention has been done to the low energy evolution and the various quark thresholds below the
top quark scale, what was also not considered in \cite{zol1,zol2}.

We now use the same numerical code to compute the scalar coupling evolution just changing the standard perturbative QCD $\beta$ function by the non-perturbative one described in Eq.(\ref{eq:x}). One very important point is to recall that our equations are in agreement with the boundary conditions given by the experimental $M_t$ and $M_H$ masses and the one of
Eq.(\ref{eq:77}). In the IR region, in the limit of two light quarks, the non-perturbative QCD $\beta$ function and the coupling generated by it is consistent with Eq.(\ref{eq:7}), which
can be considered the condition that fix our scheme, and in the ultraviolet region, with six active quarks, the effect of the running dynamical gluon mass is quite small and its contribution is one order of magnitude below the uncertainty in the coupling determination at the $Z$ pole giving in Eq.(\ref{eq:77}). Finally the range of IR coupling values giving by Eq.(\ref{eq:7}) is the one that constrains our $m_g/\Lambda_{QCD}$ values, 
resulting in the shaded area shown in Fig.(\ref{fig1}) for the scalar coupling evolution. For most of the phenomenologically expected values of
the dynamical gluon mass the SM scalar coupling evolution does not cross the line $\lambda =0$ up to the Planck mass.  

We have computed the SM scalar coupling evolution with a very particular QCD resummation scheme, and, in this scheme, the scalar coupling evolution is positive up to the Planck mass
for a certain range of the parameter that determine the scheme, which is the dynamically generated gluon mass. The approach used here was considered previously in Ref.\cite{us3}, and
a similar result, but assuming a fixed point behavior generated by gravitational interaction, was obtained in Ref.\cite{wet}. There are many points still to be discussed in
this type of approach, which are related to the introduction of the non-perturbative conformal behavior into the RG equations. One have also to recall that the non-perturbative coupling 
considered here corresponds to an all order coupling in one specific sum of graphics, and to one scheme where the QCD $\Lambda$ scale
is fixed but the dynamical gluon mass may vary with $n_q$ (increasing its value \cite{mnq,us4}), leading to even larger $\lambda$ values at high energies, on the other hand this variation 
of $m_g$ with $n_q$ may also be erased by the introduction of massive physical quarks. This possibility has not been considered due to the lack of precise simulations
of the $m_g$ variation with $n_q$.
The infrared finite QCD coupling due to the existence of a dynamical gluon mass seems to be a reality at this point, and the QCD scheme introduced here may change
some results found in the literature \cite{zol1}, but improvements in the SDE solutions with better approximations
and the introduction of physical quarks can still modify the results that we described in this work.     

\section*{Acknowledgments}
This re\-search was par\-tially sup\-ported
by the Con\-se\-lho Nac. de De\-senv. Cien\-t\'{\i}fico e Tecnol\'ogico
(CNPq), by the grants 2013/22079-8 and 2013/24065-4 of Funda\c c\~ao de Am\-pa\-ro \`a Pes\-qui\-sa do
Estado de S\~ao Paulo (FA\-PES\-P) and by Coordena\c c\~ao de Aper\-fei\-\c coa\-mento
de Pessoal de N\'{\i}vel Superior (CAPES).

\begin{center}
\noindent\rule{5cm}{0.4pt}
\end{center}

\end{document}